\documentclass[11pt,twoside]{article}

\usepackage{asp2006}
\usepackage{graphicx}

\begin{document}
\title{Radio Loudness of AGNs:\\ Host Galaxy Morphology and the Spin Paradigm}  
\author{\L ukasz Stawarz$^{1, \, 2}$, Marek Sikora$^{3, \, 4}$, and Jean-Pierre Lasota$^{4, \, 2}$}   
\affil{$^1$Kavli Institute for Particle Astrophysics and Cosmology, Stanford University,
Stanford CA 94305, USA\\
$^2$Astronomical Observatory, Jagiellonian University, ul. Orla 171, 30-244 Krak\'ow, Poland\\
$^3$Nicolaus Copernicus Astronomical Center, Bartycka 18, 00-716
Warsaw, Poland\\
$^4$Institut d'Astrophysique de Paris, UMR 7095 CNRS, Universit\'e
Pierre et Marie Curie, 98bis Bd Arago, 75014 Paris, France}    

\begin{abstract} 
We investigate how the total radio luminosity of AGN-powered
radio sources depends on their accretion luminosity and the central black
hole mass. We find that AGNs form two distinct and well separated
sequences on the radio-loudness --- Eddington-ratio plane. We argue that
these sequences mark the real upper bounds of radio-loudness of two
distinct populations of AGNs: those hosted respectively by elliptical and
disk galaxies. Both sequences show the same dependence of the
radio-loudness on the Eddington ratio (an increase with decreasing
Eddington ratio), which suggests that another parameter in addition 
to the accretion rate must play a role in determining the jet production 
efficiency in active galactic nuclei, and that this parameter is related to 
properties of the host galaxy. The revealed host-related radio dichotomy 
breaks down at high accretion rates where the dominant fraction of luminous 
quasars hosted by elliptical galaxies is radio quiet. We argue that the huge
difference between the radio-loudness reachable by 
AGNs in disc and elliptical galaxies can be explained by the scenario according 
to which the spin of a black hole determines the outflow's power, and central 
black holes can reach large spins only in early type galaxies (following major mergers), 
and not (in a statistical sense) in spiral galaxies.
\end{abstract}

\section{Inroduction}  

Some `fundamental' questions regarding jet activity of active galactic nuclei (AGNs),
addressed in Sikora et al. (2007) as summarized in this contribution, can be formulated as follows:
Is the appearence of extragalactic jet controled by interactions with the environment, 
or mainly by the `initial conditions' for the jet launching? What are these `initial conditions'?
(e.g., mass of central black hole $\mathcal{M}_{\rm BH}$, black hole spin $\mathcal{J}$
expressed in dimensional units as $a = \mathcal{J}/\mathcal{J}_{\rm max} = c \, \mathcal{J}/G \, 
\mathcal{M}_{\rm BH}^2$, accretion rate related to the luminosity of the accreting matter $L_{\rm acc}$, etc.) 
How do the initial conditions relate to the parameters of host galaxies? How many types of jets are there? 
Is there the same physics behind all the variety? i.e., are the jets formed, accelerated and collimated 
by the same processes in all jetted AGNs? Why the efficiency of jet production is so different among 
objects very similar in all other aspects?

Only recently one could attempt to answer such questions,
because only recently we have learned how, for a large sample of substantially different AGNs, 
to estimate masses of central supermassive black holes (SMBHs; e.g., Vestergaard 2002, Woo \& Urry 2002, 
Cao \& Rawlings 2004), 
to extract accretion-related luminosities from the starlight (e.g., Ho \& Peng 2001, Ho 2002, Kharb \& Shastri 2004), 
to determine precisely morphology of host galaxies (e.g., Bahcall et al. 1997, Malkan et al. 1998, Martel et al. 1999), 
or finally how to measure weak jet-related radio emission of low-power or `radio-quiet' AGNs 
(e.g., Ho \& Peng 2001, Ho 2002, Greene et al. 2006).

\section{The Sample}   

In order to address the questions provided above, one should compare the main jet parameter, 
i.e. bulk kinetic power $L_{\rm j}$, with the main \emph{observed} parameters of the central engine, namely 
$\mathcal{M}_{\rm BH}$ and $L_{\rm acc}$, for AGNs covering many decades in radio and accretion disk luminosities. 
Hence, the resulting sample has to be \emph{by definition heterogeneous and incomplete}. In Sikora et al. (2007)
we selected sources for which:
\begin{itemize} 
\item the optical flux of the unresolved nucleus is known;
\item the total radio flux is known (includng extended emission);
\item the black hole mass can be estimated.
\end{itemize}
In order to avoid complications due to signifcant beaming and obscuration, blazar sources 
(optically violent variable quasars, highly polarized quasars, flat-spectrum radio
quasars, BL Lac objects) as well as type-2 AGNs (narrow-line radio galaxies, Seyfert 2 
galaxies) were excluded.

The remaining selected sources were divided into five subgroups: 
broad-line radio galaxies (BLRGs), broad-line radio quasars (BLRQs), Seyfert 1 galaxies 
and low-ionization nuclear emission-line region objects (Sy1s + LINERS), FR I radio galaxies
(FR Is), and finally optically-selected Palomar-Green quasars (PG QSOs). For these, the following 
parameters were determined:
\begin{itemize} 
\item nuclear $B$-band luminosity of the accretion disk $L_B \equiv \nu_B \times L_{\nu_B}$, where
$\lambda_B = 4400$\,\AA\, and, by assumption, bolometric accretion-related luminosity 
$L_{\rm acc} \approx 10 \times L_B$;
\item total jet-related radio luminosity $L_R \equiv \nu_R \times L_{\nu_R}$, where $\nu_R = 5$\,GHz
and, by assumption, the jet kinetic luminosity $L_{\rm j} \propto L_R$;
\item radio-loudness parameter $\mathcal{R} \equiv L_{\nu_R}/L_{\nu_B} \approx 10^5 \times (L_R/L_B)$; 
\item accretion luminosity $\lambda \equiv L_{\rm acc}/L_{\rm Edd} \approx 10 \times (L_B/L_{\rm Edd})$, where
the Eddington luminosity for a given mass of a black hole is 
$L_{\rm Edd} = 4 \pi \, G \mathcal{M}_{\rm BH} \, m_p c / \sigma_{\rm T} 
\approx 10^{38} \times (\mathcal{M}_{\rm BH}/M_{\odot})$\,erg\,s$^{-1}$.
\end{itemize}
The details of the sample selection and the appropriate references are provided in Sikora et al. (2007).

\section{Observational Facts}   

As shown in figure 1, different considered types of AGNs form two clear sequences on the $L_R-L_B$ plane.
The upper sequence is almost exclusively populated by objects hosted by elliptical galaxies with 
$\mathcal{M}_{\rm BH} \geq 10^8 \, \mathcal{M}_{\odot}$. The lower sequence is populated by AGNs 
hosted by both elliptical and disk galaxies. Both sequences show however similar behavior 
$L_R \propto L_B^n$ with $0<n<1$ at low $L_B$, and some kind of plateau $L_R \propto const$ at high $L_B$.
Interestingly, the same two sequences emerge also on the $(L_B/L_{\rm Edd})-(L_R/L_{\rm Edd})$ plane, as shown
in figure 2.

Elliptical-hosted and spiral-hosted AGNs remain well separated also on the $\mathcal{R}-\lambda$ plane, as shown
in figure 3.
It can be noted that the radio loudness parameter $\mathcal{R}$ increases with decreasing Eddington ratio 
$\lambda$, as previously pointed out by Ho (2002). In addition to this, some saturation of radio 
loudness at low accretion rates $\lambda < 10^{-3}$ can be observed. 
The above trend is followed separately by the two upper and lower sequences, 
which we call hereafter `radio-loud' and `radio-quiet', respectively. Note, 
however, that with the standard criterion of radioloudness ($\mathcal{R} > 10$, Kellerman et al. 1989), all
the low-power, spiral-hosted AGNs characterized by low accretion rates would be rather classifed as `radio-loud'
(see in this context Ho \& Peng 2001, Ho 2002).

Finally, one can observe that within the considered sample 
AGNs with $\mathcal{M}_{\rm BH} > 10^8 \, \mathcal{M}_{\odot}$ 
seem to reach values of radio loudnes $>1000$ times larger than AGNs with $\mathcal{M}_{\rm BH} < 
10^8 \, \mathcal{M}_{\odot}$, as also noted previously by McLure \& Jarvis (2004). This is shown in
figure 4.

Interestingly, in the case of jetted X-ray binaries (XRBs), at low accretion rates radio luminosity scales 
with accretion (X-ray) luminosity like $L_R \propto L_X^{0.7}$ (Gallo et al. 2003). At high accretion rates, 
$\lambda \geq 0.01$, such scaling breaks down, and the jet production starts to be highly intermittent. 
Similarity of this behavior to the case of elliptical-hosted AGNs considered here
--- where for low accretion rates $L_R$ is a monotonic function of $L_{\rm acc}$, while at high accretion 
rates bimodial distribution of $L_R$ suggests highly intermittent jet production efficiency ---
is striking (see figure 2). This was discussed previously by, e.g., Merloni et al. (2003).
However, $L_R \propto L_{\rm acc}^n$ at low $\lambda$ trend is followed separately by the both `radio-loud' 
and `radio-quiet' sequences in the analyzed sample. Also, jet intermittency at high accretion rates is 
restricted to AGNs hosted by ellipticals. \emph{This shows that yet another parameter in addition to the accretion 
rate must play a role in determining the jet production efficiency, and that this parameter is related to 
properties of the host galaxy.}

\begin{figure}[!ht]
\centering
\includegraphics[scale=1.1]{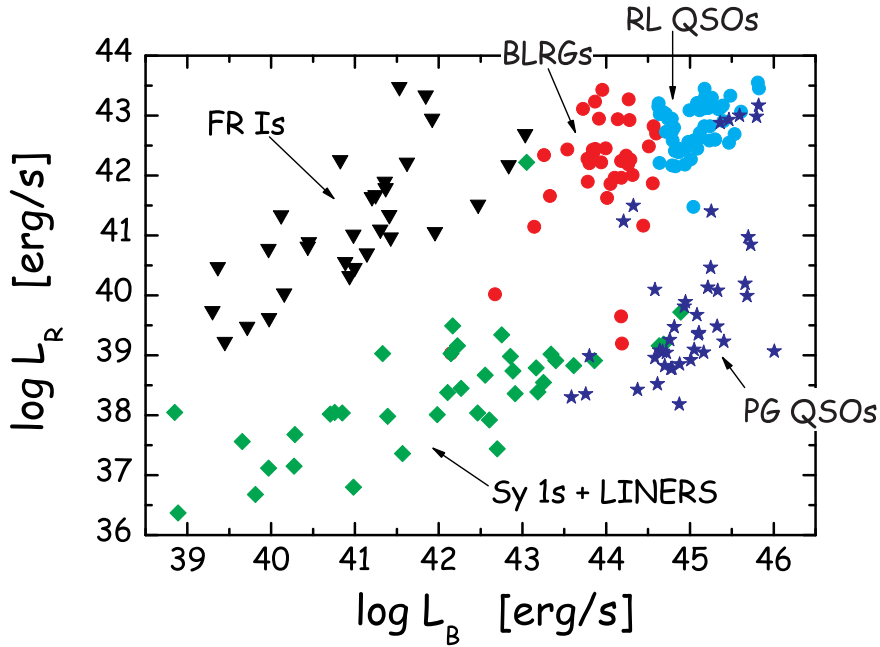}
\caption{$L_R$ vs. $L_B$ for the analyzed sample.}
\end{figure}

\begin{figure}[!ht]
\centering
\includegraphics[scale=1.1]{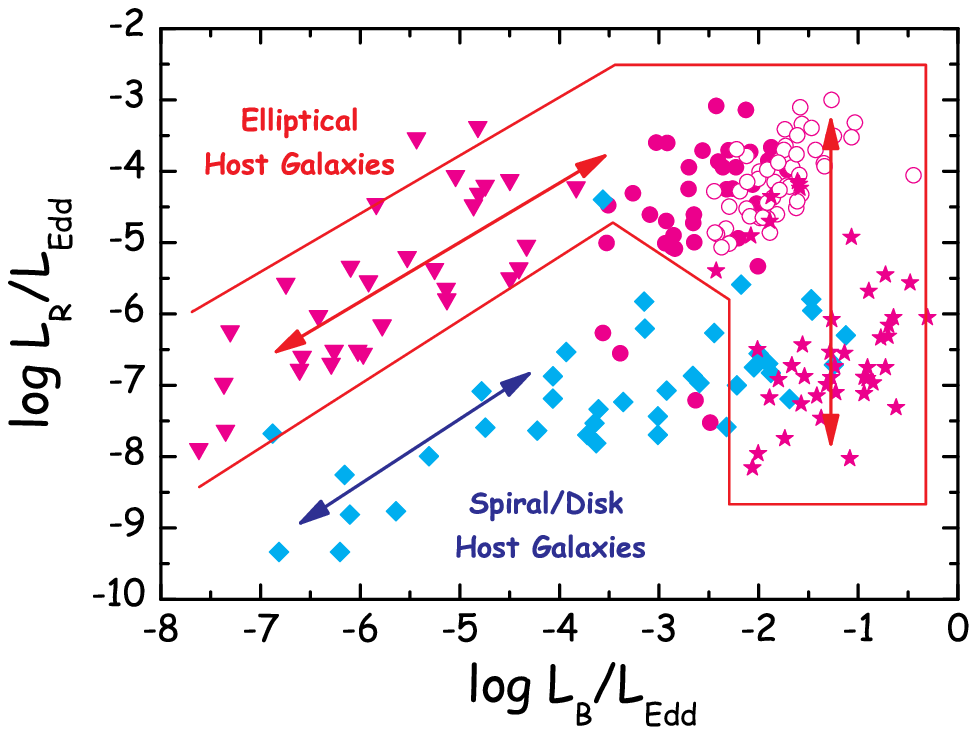}
\caption{$L_R/L_{\rm Edd}$ vs. $L_B/L_{\rm Edd}$ for the analyzed sample.}
\end{figure}

\begin{figure}[!ht]
\centering
\includegraphics[scale=1.1]{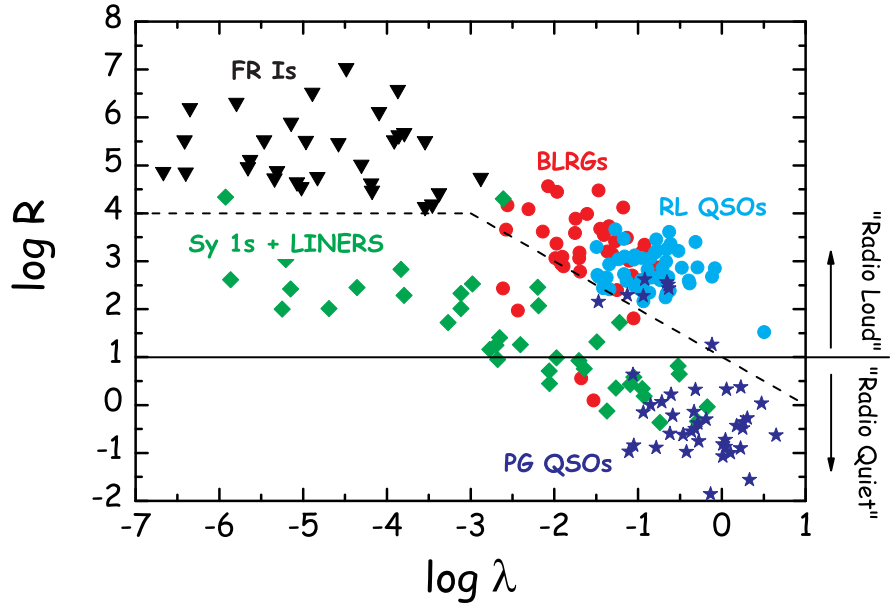}
\caption{$\mathcal{R}$ vs. $\lambda$ for the analyzed sample.}
\end{figure}

\begin{figure}[!ht]
\centering
\includegraphics[scale=1.1]{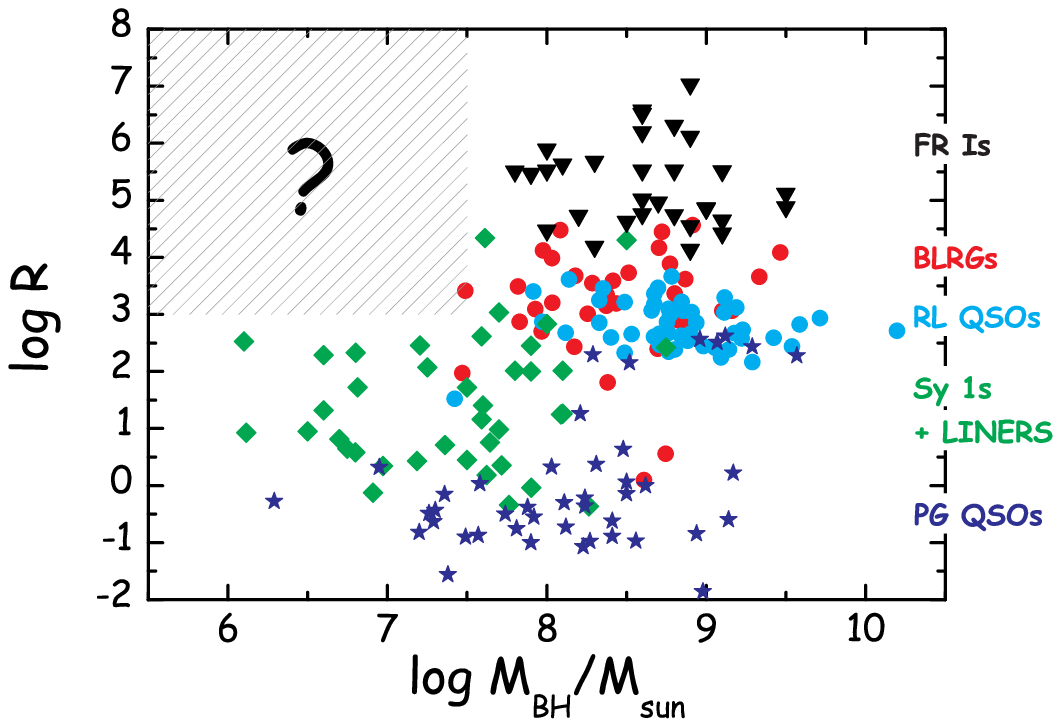}
\caption{$\mathcal{R}$ vs. $\mathcal{M}_{\rm BH}$ for the analyzed sample.}
\end{figure}

With the additional parameter we would like to therefore explain:
\begin{itemize}
\item Why the efficiency of jet production can be much larger in elliptical-hosted AGNs than in spiral-hosted AGNs?
\item Why the jet activity at high accretion rates in elliptical-hosted AGNs is highly intermittent?
\item Why there is only one `radio-loudness sequence' in XRBs?
\end{itemize}

\section{The Spin Paradigm and Some Speculations}

Is it a spin of the central SMBH which determines jet production efficiency in AGNs?
Indeed, as discussed by Blandford (1990), assuming that a jet is powered by the rotating black hole 
via the Blandford-Znajek mechanism (Blandford \& Znajek 1977), one can expect the black hole spin
playing a major role in this respect (see also Wilson \& Colbert 1995, Hughes \& Blandford 2003).
If this is the case, then one has to explain why SMBHs in ellipticals have on average much larger 
spins than SMBHs in disk-galaxies (and how the jet production for rapidly rotating SMBHs can be intermittent,
an issue which however is not addressed in this short contribution; see Sikora et al. 2007).
In other words, one has to relate the spin parameter to the morphology of the host, since the presence of 
the two revealed sequences of AGNs on the $L_{\rm j}-L_{\rm acc}$ planes are related to the type of host
galaxy. (Interestingly, the spin paradigm can account for a single `radio-loudness sequence' in XRBs, 
since the black hole spin in XRBs is not expected to evolve during the lifetime of these systems.)

In Sikora et al. (2007), we speculate that the spin evolution of SMBHs in spiral galaxies is limited by multiple accretion events with random orientation of angular momentum vectors and small increments of mass, $m \ll m_{\rm align}
\sim a \, \sqrt{R_{\rm S}/R_{\rm W}} \, \mathcal{M}_{\rm BH}$, where $R_{\rm S} = 2 \, G \mathcal{M}_{\rm BH}/c^2$ is the Schwarzschild radius and $R_{\rm W} \sim 10^4 R_{\rm S}$ is the distance of the warp produced by the Bardeen-Petterson
process in the accretion disk, which at large distances is inclined to the equatorial plane of the rotating black hole (Bardeen \& Petterson 1975). Thus, one can expect small SMBH spins in spirals, since the proposed accretion mode may spin-up but also spin-down central black holes. This picture is consistent with the observed short lifetimes of individual accretion events in Syfert galaxies (see in this context Kharb et al., 2006), and random orientation of Syfert jets relative to the axis of the host (Schmitt et al., 2001).
Unlike spiral galaxies, every elliptical underwent at least one major merger in its past, followed by the accretion of mass $m \gg m_{\rm align}$, which then will always spin-up the central SMBH. Thus, all black holes in ellipticals may spin rapidly ($a > 0.9$ if $m \sim \mathcal{M}_{\rm BH}$). This is consistent with large average spin of black holes found for quasars (So\l tan 1982).

\section{Conclusions}

The upper boundaries of radio-loudness of AGNs hosted by giant elliptical galaxies are by $\sim 3$ orders of magnitude larger than upper boundaries of radio-loudness of AGNs hosted by disc galaxies.
Both populations of spiral-hosted and elliptical-hosted AGNs show a similar but distinct dependence of the upper bounds of the radio loudness parameter on the Eddington ratio (the radio loudness increases with decreasing Eddington ratio, faster at higher accretion rates, slower at lower accretion rates).
In Sikora et al. (2007) we propose that the huge, host-morphology-related difference between the radio-loudness reachable by AGNs in disc and elliptical galaxies can be explained by the scenario according to which the spin of a black hole determines the outflow's power, and central black holes can reach large spins only in early type galaxies (following major mergers), and not (in a statistical sense) in spiral galaxies.

The above conclusions regarding observational facts (presence of the two well-separated sequences on 
the $L_{\rm j}-L_{\rm acc}$ planes in particular) were recently 
supported by Panessa et al. (2007) and Maoz (2007), while the proposed
revised spin paradigm for the jet production efficiency was investigated further by Volonteri et al. (2007).

\acknowledgements 
M.S. was partially supported by Polish MEiN grant 1 P03D 00928 and
in 2005 by a CNRS ``poste rouge" at IAP. {\L}.S. acknowledges
support by the MEiN grant 1 P03D 00329 and by the ENIGMA Network
through the grant HPRN-CT-2002-00321.
JPL was supported in part by a grant from the CNES. This research was supported in
part by the National Science Foundation under Grant No. PHY99-07949
and by the Department of Energy contract to SLAC no. DE-AC3-76SF00515.

\end{document}